**Hawking-Unruh radiation as irreversible consequence of radiative action in dynamics**

**Dasarathi. Das***
(Bhabha Atomic Research Centre
Trombay, Mumbai 400 085.

(e-mail: dasd1951@gmail.com)

Abstract

Hawking-Unruh thermal state of warm surrounding field encountered in non-inertial frames is shown to be a real phenomenon, a marker of nonstationary dynamic evolutions. In accelerated motion of a charged particle it is shown that the recoiled damping effect of Larmor radiation relaxation leads to distinctive thermal power, which is akin to that of Hawking-Unruh radiation from warm surrounding field of the accelerated charge. The damping effect from recoil-momentum of transverse electromagnetic field is worked out by considering torque imparted to the inherently existing angular evolution of spherically polarized vacuum field around the point-like charged particle in acceleration. Hawking-Unruh effects is generally noted to be a universal marker of decoherence in evolution in all scales of microscopic, macroscopic and megascopic systems. Besides detailing the case of electrodynamics, the various efficacies of H-U relaxation are considered in the nonstationary evolutions.

---

*(Presently, superannuated from the official position)

## 1.0 Introduction

Applying quantum field theory in curved space, Stephen Hawking predicted [1] that the strong gravitational force near event horizon of Schwarzschild black holes leads to the emission of a thermal radiation, which resembles black-body radiation. For black hole of mass $M$ it is shown that the thermal radiation corroborates to temperature of $T = \hbar c^3 / 8\pi\kappa_B GM$ , where $\hbar$ , $\kappa_B$ , $c$ , $G$ are respectively the Planck constant, Boltzmann constant, signal speed, and the gravitational constant. The temperature expression can be equivalently written as $T = \hbar g / 2\pi c\kappa_B$ , $g$ being gravitational acceleration at the event horizon. Black holes that are normally encountered in cosmology are having mass greater than mass of the sun and their thermal emissions correspond to temperatures that are too low to be observed within the ubiquitous microwave background of about 2.725 K. Following Hawking's prediction, William Unruh et al [2] inferred that a noninertial observer while moving with an acceleration of magnitude $\dot{\mathrm{v}}$ should observe the thermal radiation emission in the surrounding vacuum field with similar functionality, namely, $T = \hbar\dot{\mathrm{v}}/2\pi \mathrm{c}\kappa_B$ . The respective predictions are nonlocally assessed using the quantum framework applied to the curved space (Hawking's case) and to quantum field theory (Unruh's case), and applying Bogoliubov transformation in interpreting particle/photon creations from virtual states in the surrounding vacuum field near event horizon in the respective cases, namely, a black hole and noninertial frames. According to their interpretations, the quantum states of the surrounding field characteristically pose to have thermally distributed populations, the warmness of the surrounding being proportional to the acceleration characterizing the referred frame/s. To observe the thermal effect calls for generating very high accelerations. Thermal effect of 1K requires acceleration as high as $2.5 \times 10^{18} \mathrm{ms}^{-2}$. Nevertheless, confirmations of the thermal emission have been claimed recently by a number of investigators through laboratory experiments by simulating the characteristics of event horizon of a black hole in optical material [3], in water waves [4,5], and in BEC condensate [6,7], These simulated studies create analogous situations of event horizon to decipher states that are accessible and inaccessible for probing mimicking thereby the states of photon/particle pairs emitted near the black hole. By probing on the accessible states, the respective studies claim to have confirmed the thermal-like population distribution as well as distant correlation with the inaccessible states.

The predicted similarity of thermal states in Hawking radiation and in Unruh radiation is understandable by considering Einstein's equivalence principle applied to the noninertial frames including gravitational frames. Nonetheless, question remains as to whether the thermally populated vacuum states referred in the two cases could characterize the event horizon in terms of observable effects like thermal losses as per the predicted temperatures. According to Hawking's prediction, the power loss amounts to mass reduction of the concerned black hole. The reported studies with black hole analogs claim this from the observed/interpreted energy distributions of the water waves, in BEC condensates or in the transported laser pulses through extreme nonlinear fibre optics. However, question persists on whether the concerned systems defining event horizon incur thermal losses in reality according to the nonlocally assessed population distribution.

The states of spontaneous emission of photon/particle pairs from event horizon of a black hole, or, of an accelerated frame can be referred generally as critically perturbed states of nonstationary evolutions. There, one would expect that the omnipresent and incessant nonlocal mitigation to reinstate coherence in the evolutions concerned fails leading to stress yield under the critically high perturbing stress with spontaneous release of thermal radiation as indicated by the two predictions. The perturbing agency, be it gravitational field or externally applied field for the accelerated evolutions, is expected to register the relaxation loss in the mitigation process; relaxations being from gravitationally stressed vacuum field, or, from nonstationary dynamics of virtual particles/photons in vacuum field relative to noninertial observer.

The realities of thermal power losses of Hawking and Unruh can be ideally analyzed with electrodynamics of accelerated charged particle, which has been introspected in details by many investigators. Classical electrodynamics normally describes self field loss of the accelerated charge as radiation emission which is however in regular (ordered) form, namely, the Larmor radiation. In the present context, one thus introspects on whether the accelerated dynamics could register any thermal power loss by recoil effect of the regular radiation emission. Some earlier studies have designated the recoil effect as radiation damping, though damping loss is hardly substantiated. Quantum mechanical analysis using microscopic reversibility in light scattering confirms only the regular part of radiation emission from accelerated charged particle. Nevertheless, Unruh's prediction of noninertially observable thermally populated ground and excited states of vacuum field is an offshoot of quantum field theory. The predicted property of noninertial observer gives renewed interest to reanalyze for any thermal power loss from accelerated motion of a charged particle in electromagnetic field.

According to the analysis of a laboratory observer, an accelerated charge (mass $m_0$ and charge $q$) suffers recoil related displacement. Work involved therein does not explicitly figure in the power conservation equation, which shows that power spent by external field manifests as increasing kinetic energy of the charge and Larmor radiation loss. Displacement of the jerk force, $2q^2\ddot{\bar{\mathrm{v}}}/3c^3$ ($\ddot{\bar{\mathrm{v}}}$ being 3-jerk) accounts for the accelerated growth rate of kinetic energy in dynamically relevant time period of $\tau_0 \equiv 2q^2/3m_0c^3$, and for the Larmor power loss. Even though the jerk force is having finite transverse component, $\dot{\bar{\mathrm{e}}}$, ($\ddot{\bar{\mathrm{v}}} \equiv d(\dot{\mathrm{v}}\bar{\mathrm{e}})/\mathrm{dt} = \ddot{\mathrm{v}}\bar{\mathrm{e}} + \dot{\mathrm{v}}\dot{\bar{\mathrm{e}}}$, $\bar{\mathrm{e}} = \dot{\bar{\mathrm{v}}}/\dot{\mathrm{v}}$), which can be perturbed by the radiation recoil, the corresponding power loss cannot be recognized by the displacement type of work mentioned above. This is also clear from the fact that the direct component of jerk ($\ddot{\mathrm{v}}\bar{\mathrm{e}}$) on displacement generates an opposing power term to cancel out displacement effect of the transverse component ($\dot{\mathrm{v}}\dot{\bar{\mathrm{e}}}$). In displacement of direct jerk component as, $(2\mathrm{q}^2/3c^3)\ddot{\mathrm{v}}\bar{\mathrm{e}} \bullet \bar{\mathrm{v}} \equiv \tau_0\ddot{\mathrm{k}} - (2\mathrm{q}^2\dot{\mathrm{v}}^2/3c^3)$ - $(2\mathrm{q}^2/3c^3)\dot{\mathrm{v}}\dot{\bar{\mathrm{e}}} \bullet \bar{\mathrm{v}}$, ($\mathrm{k}$ being the kinetic energy) the last power term cancels out the power associated with the transverse component. However, the impact due to radiation recoil on mass $m_0$ of the charged particle generates a velocity component given by, $\bar{\mathrm{v}}_{\mathrm{recoil}} = -\dot{\bar{\Pi}}\tau_0/m_0 \equiv -(2q^2\dot{\mathrm{v}}^2\bar{\mathrm{n}}_{\mathrm{rad}}/3c^4)\tau_0/m_0$, where $\tau_0 \equiv (2q^2/3m_0c^3)$ is the time of recoil-impact, and $\dot{\bar{\Pi}}$ is the rate of Larmor radiation momentum ($\bar{\mathrm{n}}_{\mathrm{rad}}$ being unit radiation vector). It is interesting to note that the implied power loss, $\dot{\bar{\Pi}}^2\tau_0/m_0$, is having $\dot{\mathrm{v}}^4$ dependence, which is different in nature as compared to the $\dot{\mathrm{v}}^2$ dependence of Larmor radiation loss. The recoiled based power is functionally distinct from those involved in the power balance equation. This unexplored aspect of radiation recoil needs further attention as it might have underlying connection with Hawking-Unruh type radiation emission. It is important to mention here that in the case of uniformly accelerated charged particle although jerk has null expectation value ($\langle \ddot{\mathrm{v}} \rangle = 0$), the transverse force component, $(2\mathrm{q}^2/3c^3)\langle \dot{\mathrm{v}}\dot{\bar{\mathrm{e}}} \rangle \equiv \tau_0 \langle m_0 \dot{\mathrm{v}}\dot{\bar{\mathrm{e}}} \rangle$, has finite magnitude in the dynamics, which is nonstationary with the observable acceleration; finite torsion ($\dot{\bar{\mathrm{e}}}$) that results from evolution of unit acceleration vector in the critically perturbed motion by the external electromagnetic field corroborates to $\mathrm{d}\langle \bar{\mathrm{e}} \rangle/\mathrm{dt} = 0$, $\langle \mathrm{d}\bar{\mathrm{e}}/\mathrm{dt} \rangle \neq 0$. In general, $\dot{\bar{\mathrm{e}}}$ evolves as $\dot{\bar{\mathrm{e}}} = \sum_{\mathrm{k}} \mathrm{c_k}(\text{-i}\omega_\mathrm{k})\exp(\text{-i}\omega_\mathrm{k}\mathrm{t})$, where $\sum_{\mathrm{k}} \mathrm{c_k c_k^*} = 1$, and $\mathrm{c_k c_{k'}^*} = \delta_{\mathrm{kk'}}$ (Kronecker delta function) maintains unit vector property of $\bar{\mathrm{e}}$. Electrodynamics with constant expectation value of the acceleration allows its evolution entailing internal frequency characteristics related to $\tau_0^{-1}$. And the relaxation losses remain unabated in the nonstationary evolution with the observable uniform linear acceleration. Torsion, $\dot{\bar{\mathrm{e}}}$, the pertinent property in disordered relaxation gets established on the polarized vacuum field inherently existing around the point-like charged particle. The spherically polarized field having dynamically relevant radius of $r_0 = 2q^2/3m_0c^2$ experiences torque stress of $\bar{F} = \bar{\Gamma}\dot{\mathrm{e}}/c \equiv (2\mathrm{q}^2/3c^3)\dot{\mathrm{v}}\dot{\bar{\mathrm{e}}}$, ($\bar{\Gamma} \equiv [r_o\bar{\mathrm{n}}_\mathrm{r} \times m_0\bar{\mathrm{v}}]$, $\bar{\mathrm{n}}_\mathrm{r}$ being unit radial vector) in the accelerated motion. In radiation recoiling, the transverse electromagnetic momentum $-\dot{\bar{\Pi}}\bar{\mathrm{n}}_{\mathrm{rad}}$ with $\bar{\mathrm{n}}_{\mathrm{rad}} \equiv \bar{E} \times \bar{H}/EH$, ($\bar{E}$ and $\bar{H}$ being electric and magnetic field components of equal magnitudes), perturbs the torsional plane, $(\bar{\mathrm{e}}\times\bar{\mathrm{n}}_\mathrm{r})$, with $\dot{\bar{\mathrm{e}}} \equiv (\bar{\mathrm{e}}\times\bar{\mathrm{n}}_\mathrm{r})\dot{\mathrm{e}}$ of the dynamically relevant sphere; $\bar{\mathrm{n}}_\mathrm{r}$ being the unit radial vector. It can be shown by ab-initio analysis of electrodynamics with the consideration of nonlocal mitigation on external perturbation that the transverse component of jerk representing the torque force $\bar{\Gamma}(\dot{\mathrm{e}}/c)$, dynamically counterbalances the electromagnetic momentum recoil as $\langle \bar{\Gamma}(\dot{\mathrm{e}}/\mathrm{c}) - \dot{\bar{\Pi}} \rangle_{cr} = 0$, $\dot{\bar{\Pi}} \equiv 2q^2\langle \dot{\mathrm{v}}^2 \rangle \bar{\mathrm{n}}_{\mathrm{rad}}/3c^4$; symbol, $\langle\ \rangle_{cr}$ designates expectation values at the critical perturbation. Salient features of the analysis, which is essentially based on Lagrangian formalism but provides an abridged version of classical and quantum mechanics while optimizing radiation loss/gain in delocalized dynamic passage, are described elsewhere (appendix in reference, [8]). The abridged version could be extended to the case of describing entagled geodesics in quantized space fabrics, which is described in this text (subsection 3.1.1). In this version the field-particle interaction are influenced by the nonlocal mediation characteristics which have been described without resorting to the conventional approach of axiomatic introduction of quantum features. The mediation incessantly preserves unitarity in field-particle evolution, and the virtual counteraction of the nonlocal field turns into a real one in the form of the recoil effect from the relaxing radiation at the criticality. The analysis shows that there is angular displacement rate under the perturbing torsional stress at criticality, which is expressible as $\langle \dot{\bar{\mathrm{e}}} \rangle_{cr} = \langle \dot{\mathrm{v}}/\mathrm{c} \rangle_{cr} \bar{\mathrm{n}}_{\mathrm{rad}}$, and concludes that polarized vacuum field around the accelerated charge being torsionally perturbed relieves off the torque stress as thermal radiation. The thermal power loss is additional over the Larmor radiation, and it is borne by the external field while sustaining the acceleration. The thermal relaxation analysis part is discussed in the next section.

### 2.0 Thermal radiation from torsional stress in radiation recoil

The radiation recoil leads to driven damped oscillation of internal modes of the polarized sphere of vacuum field with the dynamically relevant radius of $(2q^2/3m_0c^2)$ existing around point charge. In a recoil event, two of the three mutually orthogonal linear harmonic modes of the spherical oscillator forming the canonically averaged plane, $\overline{\mathrm{n}}_{\mathrm{rad}} = \langle \overline{\mathrm{n}}_r \times \overline{\mathrm{e}} \rangle_{cr}$, ($\overline{\mathrm{n}}_r$ being unit radius vector over the polarized sphere) are torsionally driven under the transverse electromagnetic force, $-\dot{\overline{\Pi}}$. Using standard analytical results of driven damped oscillation [9] under the critical perturbation one finds that energy relaxation occurs at the rate of $(2\varsigma m_0\omega_0)^{-1}\Pi^2\zeta = \dot{\mathrm{E}}_{\mathrm{relax}}(\mathrm{say})$. $\varsigma$ is damping parameter, which is unity for the critically damped oscillation occurring with the fundamental angular frequency $\omega_0$ of the linear harmonics. Noting that one full oscillation involves sweeping over four times the radius, one expresses $\omega_0$ as $\omega_0 = 2\pi/4\tau_0 = 2.5\times10^{23}$ s$^{-1}$, ($\tau_0 = 2q^2/3m_0c^3$ ~6.3x10$^{-24}$s). In the relaxation expression, the multiplier, $\zeta$ represents recoil driven oscillation factor as $\zeta = [(2\Delta\omega/\omega_0)^2+1]^{-1}$, where $\Delta\omega = \omega_0 - \omega_{\mathrm{rad}}$, $\omega_{\mathrm{rad}}$ representing frequency of recoiling photons. $\Delta\omega << \omega_0$ represents the near resonance cases, for which $\zeta \sim 1$, whereas $\Delta\omega \sim \omega_0$ represents off resonance cases for which $\zeta \sim 1/5$. $\dot{\mathrm{E}}_{\mathrm{relax}}$ has finite magnitude in nonstationary evolution having nonzero expectation value of acceleration, ($\langle \dot{\mathrm{v}}^2 \rangle \neq 0$). The power loss in critical damping in fact can be rewritten as $\dot{E}_{\mathrm{relax}} = [(\dot{\overline{\Pi}} \bullet \delta\overline{\Pi}/m_0)/\pi]\zeta$, where, $\dot{\overline{\Pi}} = (\dot{E}_{\mathrm{Larmor}}/c)\overline{\mathrm{n}}_{\mathrm{rad}} \equiv (m_0c/\tau_0)(\dot{\mathrm{v}}/\dot{\mathrm{v}}_{\mathrm{max}})^2\overline{\mathrm{n}}_{\mathrm{rad}}$, and $\delta\overline{\Pi}/m_0 = \dot{\overline{\Pi}}\tau_0/m_0$, the recoil velocity, and $\dot{\mathrm{v}}_{\mathrm{max}} = c/\tau_0 \equiv 3m_0c^4/2q^2$. The limiting acceleration $\dot{\mathrm{v}}_{\mathrm{max}}$ arises since, at and beyond this acceleration limit, the regular (Larmor) relaxation loss will be at the expense of proper energy of the accelerated particle itself. Within the limit of $\dot{\mathrm{v}}_{\mathrm{max}} \geq \dot{\mathrm{v}} \geq 0$ the particle can dynamically evolve under the critical perturbations without any loss of its self energy; relaxation power is derived from the perturbing external field. As compared to Larmor radiation loss, $\dot{E}_{\mathrm{relax}}$ is insignificant under normal acceleration values as can be seen by the ratio: $\dot{E}_{\mathrm{relax}}/\dot{E}_{\mathrm{Larmor}} = \zeta[(\delta\Pi/m_0c)/\pi] = \zeta\left(\dot{\mathrm{v}}^2/\dot{\mathrm{v}}_{\mathrm{max}}^2\right)/\pi$, $1 \geq \zeta \geq 1/5$. At the limiting acceleration, $\dot{\mathrm{v}}_{\mathrm{max}}$, polarized sphere is recoiled predominantly with radiations of near resonant frequencies, $\omega_0 \sim \dot{\mathrm{v}}_{\mathrm{max}}/c$ ($\omega_0 \approx 1.6\times10^{23}$ s$^{-1}$) attaining thereby the maximum value of $\dot{E}_{\mathrm{relax}}/\dot{E}_{\mathrm{Larmor}} = 1/\pi$.

It can be seen that the thermal power in torsionally driven damping is expressible as $\dot{E}_{\text{relax}} = \dot{E}_{\text{Larmor}}\zeta\left(\dot{\text{v}}^2 / \dot{\text{v}}^2_{\max}\right)/\pi \equiv \zeta[S_0\alpha\hbar\dot{\text{v}}^4 / 6\pi^2c^6]$, where $\alpha \equiv q^2 / c\hbar \sim 1/137$ is the fine structure constant, and $S_0\,(=4\pi c^2\tau_0^2)$ is the surface area of the dynamically significant sphere of radius, $c\tau_0$. $\dot{E}_{\text{relax}}$ has the bi-quadratic functionality in acceleration indeed. Considering that the relaxation results in thermal radiation emission uniformly over the entire $4\pi$ solid angle of the polarized sphere, the power density expression can be rearranged as $(\dot{E}_{relax} / S_0) = [\zeta(160\alpha)][\sigma(\hbar\dot{\text{v}} / 2\pi c\kappa_B)^4]$, where $\sigma = \pi^2\kappa_B^4 / 60\hbar^3c^2$, is the Stefan-Boltzmann constant. The power density corroborates to black body emission temperature of $\text{T}_{\text{rad}}(K) = (160\alpha\zeta)^{1/4}[\hbar\dot{\text{v}} / 2\pi c\kappa_{\text{B}}] \approx 1.04\zeta^{1/4}[\hbar\dot{\text{v}} / 2\pi c\kappa_{\text{B}}]$, where, $\kappa_{\text{B}}$ is Boltzmann constant. $\zeta$ assumes unity at the highest possible nonlocal defense, which is at the critical acceleration of $\dot{\text{v}}_{\max}$. With lowering of acceleration, $\zeta$ reduces and converges to the lowest limit of 1/5, $(1/5 \le \zeta \le 1)$. Nonetheless, $\text{T}_{\text{rad}}$ has semblance to Hawking-Unruh radiation temperature [1,2].

The above analysis thus shows that an accelerated charged particle does have thermal relaxation due to recoiled damping from Larmor radiation momentum. The thermal power is having bi-quadratic dependence on acceleration as in Hawking-Unruh radiation, which a noninertial observer commoving with the accelerated charge, assigns it to be the emission characteristics of surrounding vacuum field exhibiting thermally elevated state. Relative to the noninertial observer, virtual particles/photons in vacuum field are under acceleration and result in the radiation emission in the nonstationary state. Observer of lab frame in contrast notes it as dissipative power loss from torsionally damped oscillations of spherically polarized vacuum field inherently associated with the charge executing accelerated motion. And he infers that thermal relaxation loss from the spherically polarized surface of the charge is borne by the external supply used for sustaining the accelerated motion; the supply bears all the relaxation losses, regular (Larmor) and disordered ones with their proportions of $\dot{E}_{\text{relax}}/\dot{E}_{\text{Larmor}} = \eta = \zeta(\dot{\text{v}}^2 / \dot{\text{v}}^2_{\max})/\pi$, $0 \le \eta \le 1$.

As exemplified in the accelerated motion of a charged particle, nonstationary dynamics in general belong to the critically perturbed state of evolutions where both ordered and disordered relaxations occur. In the wake of reaching the criticality where perturbation starts superseding the virtual counteraction limit of the incessantly mitigating nonlocal field the relaxation signatures show up. The criticality at a given acceleration is generally defined by the torsional perturbation as expressed by angular acceleration in the case of electrodynamics as $(\dot{\text{v}}/c)^2 \equiv \dot{\text{e}}^2_{cr}$, where jerk manifested torque is counterbalanced by maximum defense of the mitigating nonlocal field; defense limit at a given acceleration is achieved by compromising the mediating speed ($\text{w}$; $c \le w \le \infty$) as $\lim_{\text{w}\to c}(\dot{\text{v}}^2/\text{w}^2) \equiv \dot{\text{e}}^2$, $\overline{\text{w}}$ being the phase velocity of the nonlocally mediated evolution (appendix in reference, [8]). As the uprising perturbation approaches the critical value, the nonlocal field to tackle perturbation compromises on its wave communication speed $(\text{w})$ until it attains the lowest limit equal to the signal speed. The criticality thus defines the parametric boundary for nonlocal defense limit; within the boundary, nonlocal field could defend coherent evolution in virtual way successfully, and on the boundary the defensive reaction leads to power relaxation at the expense of critically superseding perturbing field.

Decoherence characteristics at critical perturbation, where H-U thermal effect constitutes the disordered part of the overall relaxation are evidenced not only in microscopic evolution but also in the evolutions of macroscopic and megascopic systems. In dynamic frictional phenomenon, the relaxations at decoherence of oscillating surface phonon modes of two mutually rubbing solid surfaces involve loss of regular sonic power besides frictional damping loss. Taking similitude of the two types of relaxations noted in electrodynamics, it is possible to quantify precisely the frictional coefficients for metals above and below its superconducting transition. Frictional properties of metals above and below the superconducting transitions are elaborated elsewhere [8]. As for the megascopic case, the observed space expansion characteristics can be recognized and quantified by vacuum field mitigated nonstationary evolution, where the relaxation features in decoherence can be understood as the unseen (dark) energy/matter manifesting from the mitigating field, which is characteristically possessing two distinguishable energy states in quantifiable proportions; proportions are shown corroborating to the reported contents of matter and energy [10] in the gravitating space. The universal decoherence characteristics apparently provides the scenario of evolutionary beginning of the gravitating space. Salient features are discussed in next sections.

**3.0 H-U state as a specific feature of decoherence in nonstationary space evolution**

Manifested matters and energies in the ever expansive space signify the overall aspect of field mitigated stress relaxation features in decoherence of nonstationary evolution of gravitationally active space. Unseen energy and matter, that respectively cause space to expand against gravitational pull of matters and to facilitate structural segregation of the observable matters in localized forms within the expanding space, are the mitigated virtues of ordered and disordered relaxations. The characteristically disordered manifestation of H-U thermal state as a relaxation signature in decoherence shows up here in the form of unknown gravitationally active feature called dark matter. The proportion of the disordered relaxation in the mitigated space evolution is worked out by considering the characteristics of the mighty mitigating field as well as the observed space expansion.

3.1 <u>H-U states explored in stress relaxations of critically mitigated evolution</u>:

*(i) Relaxing stress evaluation from field mitigation characteristics*

Disordered as well as ordered type of relaxation stresses of the nonstationary evolving space can be obtained by considering the universal rule of the nonlocal mitigation that its reaction to perturbation scales down as per the acceleration characteristics $\dot{\mathrm{v}}^2/c^2$ as was noted in electrodynamics. The maximum possible value of the acceleration corroborates to that of the vacuum field, which is empirically represented by $\dot{\mathrm{v}}_{\max} \sim (c^2/\lambdabar)$. Parameter $\lambdabar$ represents the basic length of quantized space fabric which is describable with the network of entangled geodesics having dispersion property as $(2\pi/\lambdabar)^2 = k^\alpha k_\alpha \equiv (\omega/c)^2 - k^2$. (Ab-initio analysis elaborates on the dispersion characteristics (refer section 3.1.1) that supports field free evolution of test particles and defines the field energy in quantized space as $2\pi c\hbar/\lambdabar$ and correspondingly, energy density of the geodesic network as $\rho_{vac} = 2\pi c\hbar/\lambdabar^4$). Now, the evolution of geodesic distance between a pair of arbitrarily chosen space points can be generally written as $\delta R/R = (\dot{R}/R)\delta t + (\ddot{R}/R)(\delta t)^2/2 + (\dddot{R}/R)(\delta t)^3/6 + \ldots$, where $\delta\bar{R}/R$ represents the angular drift of the geodesic distance in a small time increment of $\delta t$, ($R$ being the radial length of the chosen space points from observer), and the angularly defined quantities $(\dot{R}/R)$, $(\ddot{R}/R)$, $(\dddot{R}/R)$, etc. respectively represent space expansion speed, acceleration, jerk, etc. Over and above the well established first order expansion rate of $(\dot{R}/R)$, if higher order kinematic parameters are present then these do contribute to the cumulative angular displacement between space points taken over billions of light years long space evolution history. Besides the well established Hubble's expansion parameter of this epoch as $H_0$= 70($\pm 3$) kms$^{-1}$Mpc$^{-1}$, which corroborates to the angular speed of $(\dot{R}/R)_{\text{present}} = 2.19\text{x}10^{-18}$ radian per second, the presence of acceleration in the expansion is already an experimentally confirmed fact. The acceleration, $(\ddot{R}/R)_{\text{present}}$ of the space expansion can be generally represented by square of the angular speed as $\varepsilon H_0^2$, where $\varepsilon$ is a parameter that takes care of possible deviation of acceleration on geodesic path as compared to that of the idealized radial path of $\bar{R}$. (In gravitationally active space the geodesic path deviates from the idealized circular path). Now, referring to energy-momentum stress kinematics of matter/energy and affine space curvature as described in general relativity, the space expansion characteristics (speed and acceleration values) are describable in terms of the gravitationally active density. For the observable flat space, these can be described by using Friedmann-Lemaitre-Robertson-Walker metric as

$$(\dot{R}/R)^2_{present} \equiv H_0^2 = (8\pi G/3c^2)(\rho_M + \rho_{DE}) \qquad \text{(A)}$$

$$(\ddot{R}/R)_{\text{present}} \equiv \varepsilon H_0^2 = -(4\pi G/3c^2)[(\rho_M + \rho_{DE}) + 3(p_M + p_{DE})] \qquad \text{(B).}$$

In eq.(A) and (B), $\rho_M$, and $\rho_{DE}$ are energy densities respectively of total gravitating matters (normal + dark matters), and dark energy; $p_M$ and $p_{DE}$ are the corresponding pressures, and $G$, the gravitational constant. Noting that the quantity $(8\pi G/3c^2)^{-1}H_0^2$ characterizes the critical density, $\rho_c$ of the flat space and then considering the state properties, namely, $w_M \equiv p_M/\rho_M \approx 0$, and $w_{DE} = -1$, equations (A) and (B) can be rewritten respectively as $\rho_c = (\rho_M + \rho_{DE})$, and $2\varepsilon\rho_c = (-\rho_M + 2\rho_{DE})$. The rewritten equations therefore imply the equality, $\rho_{DE}/\rho_c \equiv \Omega_{DE} = (2\varepsilon + 1)/3$. The proportions of matters and field energy in gravitationally active space are dependent on the parameter, $\varepsilon$.

It is to be noted that for $\varepsilon = 1$, the dark energy poses to be the sole constituent, which is applicable for pure vacuum field, but not so for the gravitationally active space. For the vacuum field, evolution follows stress energy equation, $(8\pi G/3c^2)\rho_{vac} = c^2/\bar{\lambda}^2$. The energy evolution of the field also represents its momentum evolution as can be verified by considering eq.(B) in its particular form, where the LHS and RHS terms respectively involving acceleration and densities/pressures are to be replaced by those of the vacuum field in which case, $\rho_M = p_M = 0$, and the acceleration and density are respectively $c^2/\bar{\lambda}^2$ and $\rho_{vac}$ with $p_{vac}/\rho_{vac} \equiv w = -1$. The unique representation of energy momentum evolution with the equation of state, $w$, shows that pure vacuum field evolves with the dark energy characteristics, $\rho_{vac} \equiv \rho^0_{DE}(\text{say}) = 2\pi c\hbar/\bar{\lambda}^4$, ($\rho^0_{DE}$ being the highest possible value of $\rho_{DE}$). Thus, the dark energy is essentially constituted of energy quanta ($c\hbar/\bar{\lambda}$) of the nonlocally mediated space fabrics. Dark energy endowed with the equation of state is rather to be taken as the highest expectation value of vacuum energy in mitigating critically highest perturbations in defending coherent states.

Evolution of mitigating field as stated above helps estimating the parameter when one makes use of the fact that the magnitude of acceleration of the gravitating space ($\varepsilon H_0^2$) is the scaled down value of the pure vacuum field, that is, $\varepsilon H_0^2 = \chi(c^2 / \bar{\lambda}^2)$, $\chi$ being scaling parameter. Noting the fact that the scaling operation must be carried out uniquely obeying energy-momentum conservation, the scaling factor connects densities of the gravitating space and mitigating field space as $\chi = \rho_c / \rho_{vac} = \varepsilon H_0^2 / (c^2 / \bar{\lambda}^2)$. Thus, one gets the equality $\varepsilon = [(c^2 / \bar{\lambda}^2) / H_0^2](\rho_c / \rho_{vac})$. In the parametric expression, it is to be noted that $H_0$ and therefore $\rho_c = (8\pi G / 3c^2)^{-1} H_0^2$ are based on observational facts of evolution properties of the cosmic sphere, whereas the field acceleration, $(c^2 / \bar{\lambda}^2)$ and the density $\rho_{vac} \equiv 2\pi c\hbar / \bar{\lambda}^4$ are derived by using the quantized dimension $\bar{\lambda}$ of cubic (virtual) lattices constituting quantized fabrics of flat space. The parameter to be estimated in consistence with the observable data, it is necessary to rewrite the relevant expression compatible with the spherical space geometry such that the basic volume $\bar{\lambda}^3$ as well as density $\rho_{vac}$ remains unaltered. The field acceleration however needs normalization by replacing the dimension, $\bar{\lambda}$ as $\bar{\lambda}' = (3/4\pi)^{1/3}\bar{\lambda}$. Considering, this normalized dimension the parameter $\varepsilon$ is replaced by $\varepsilon'$ as $\varepsilon' = [(c^2 / \bar{\lambda}'^2) / H_0^2](\rho_c / \rho_{vac})$. The rewritten parametric expression is applicable to the observable space. The corresponding field evolution with the unaltered $\rho_{vac} \equiv 2\pi c\hbar / \bar{\lambda}^4$ is given by $(8\pi G / 3c^2)\rho_{vac} = (c^2 / \bar{\lambda}'^2)$. The evolution corroborates to $\bar{\lambda}' \equiv (3/4\pi)^{1/3}\bar{\lambda} = (3/4\pi)^{1/6}(4\pi G\hbar / c^3)^{1/2}$. In the literature, the term $(4\pi G\hbar / c^3)^{1/2}$ is generally referred as Lorentz-Heaviside version of Planck's length. Now, in the normalized expression of the parameter, ($\varepsilon'$) one can substitute the values of the involved quantities, which leads to $\varepsilon' \approx 0.6213$. Using this $\varepsilon'$ value, the proportion of dark energy works out as $\Omega_{DE} = (2\varepsilon' + 1)/3 = 0.7475$. This result of about 75% dark energy composition of universe is somewhat higher than the observationally obtained value of 68.3% [10]. Noting that $\varepsilon'$ depends on the Hubbles parameter, which has been known with an uncertainty as $H_0 = 70 \pm 3$ kms$^{-1}$Mpc$^{-1}$, the estimated proportion of dark energy can be stated as $75(\pm 6)$%. Within the uncertainty, this estimation covers the observed data; for example, $H_0 = 73$ kms$^{-1}$Mpc$^{-1}$ corroborates to about 69% energy content.

This analysis made, by considering stress kinematics in the gravitating space and using the scaling law of energy-momentum of vacuum field in its critical mitigation against gravitational perturbation in the space evolution down to that existing at critical density ($\rho_c$), shows that the ultrahigh density $\rho_{vac}$ does not pose cosmological issue. Rather, it stands as testimony of the universal mitigation property of the mighty vacuum field; ultrahigh field density is endorsed by the quantum field theory. That the mitigation is scaled down according to the angular acceleration $(\dot{v}^2 / c^2)$ of nonstationary evolution was seen in the relaxation characteristics of electrodynamics. It will be seen in subsequent sections that the vacuum field mitigation endorses the space expansion property and also the classically described energy-momentum stress kinematics of gravitating matter and affine space.

*(ii) Relaxation stresses using thermodynamic modeling of space expansion characteristics*

Considering that the vacuum field mitigates on gravitationally perturbed space evolution with scaled down values of the field's energy and momentum densities, the mitigation process can be modeled as transfer of quantized states of the field to the gravitationally active space; transferred states thereby metamorphically transforming into gravitating energy/matter. The transfer takes place at an appropriate flow rate maintaining consistency in energy-momentum stresses among gravitating energy/matter in the evolving space. Recalling that dark energy and dark matter are relaxation products of the field mitigation, the dark energy density in the active space remains within the limit, $\rho_{DE} < \rho_{vac}$; $\rho_{vac} \equiv \rho^{0}_{DE}$ (say) representing the expectation value of the field energy density corroborating to the stress kinematics of $(8\pi G/3c^2)\rho_{vac} = (c^2/\bar{\lambda}'^2)$. Now, the stress transfer process can be thermodynamically described as spontaneous flow of enthalpy from the space of high pressure ($\sim \rho_{vac}$) to the relatively lower pressure region; and ideally, the flow occurs isothermally. The flow results in pressure($p$)-volume($V$) effects over and above internal energy change along with increase in kinetic energy of the gravitating matters. As will be shown now, the flow process is reflected in the space expansion characteristics that are confirmed to have speed ($\dot{R}/R \equiv H_t$, the Hubble's parameter) and acceleration ($\ddot{R}/R \equiv \varepsilon' H_t^2$); under the pressure evolution if jerk manifests, the jerk is assignable as $\dddot{R}/R \equiv \varepsilon' H_t^3$. Exploration of energy aspects of the expansion helps evaluating relative proportions of dark energy and dark matter. The proportions follow from volume expansion characteristics of the homogeneously and isotropically considered space of the observable cosmic sphere. Recalling the characteristics of geodesic distance evolution between two heavenly bodies as $\delta R = R(t)[H_t \delta t + H_t^2(\delta t)^2/2 + H_t^3(\delta t)^3/6 + ...]$, the volume expansion rate at an instant can be written as $V(t)\{\lim_{\delta t \to 0}[(1 + H_t \delta t + ...)^3 - 1]\}/\delta t \approx 3V(t)H_t$, where $V(t) = 4\pi R^3/3$. When this expansion takes place under pressure $p(t)$, the cumulative pressure-volume type work involved in the whole space evolution history is expressible as $E_{\mathrm{w}} = 3\int_{t=0}^{\tau_{\mathrm{age}}} p(t)V(t)H_t dt$, ($\tau_{\mathrm{age}}$ being the evolution time of 13.8 billion light years). Cumulatively, there is parallel accumulation of elastic energy as $E_{\mathrm{elast}} = \int_{t=0}^{\tau_{\mathrm{age}}} V(t)\dot{p}(t)dt$. The $p - V$ effects in the spontaneous flow of field energy thus leads to stress manifestations with and without the work signatures; in the elemental transfer of the enthalpy these are respectively represented by $p\delta V$ and $V\delta p$. The stress manifestation as volume displacement in the expansion of gravitationally active space can be endowed with the dark energy manifestation and the $V\delta p$ type elastic stress accumulation at constant volume can be assigned to gravitating dark matter. The manifestated stresses in the space expansion respectively with and without work signatures endorse the universal relaxation characteristics under the critical mitigation of nonstationary evolution.

Vacuum field with its very high internal pressure ($\sim \rho_{vac}$) pushes quantized cell volumes ($\bar{\lambda}^3$) one after another and thereby accomplishes the volume displacement work in the gravitating space. In the enthalpy flow process if one considers that the quantized volume units $\bar{\lambda}^3$ are transferred in synchrony with the space expansion characteristics, the pressure thrust on a unit cell proportional to $p(t)\bar{\lambda}^2$ generates the inertial force, $[(\rho_{vac}/c^2)\bar{\lambda}^3]\ddot{a}(t)$, where $(\ddot{R}/R) \approx (\ddot{a}/a)$, $a$ being the scale factor associated with the acceleration of cell being transferred. Thus, one writes, $p \propto [\rho_{vac}(\bar{\lambda}/c^2)a](\ddot{a}/a)$. Furthermore, when one considers the presence of finite jerk along with the acceleration, the jerk $(\dddot{a}/a) \approx (\dddot{R}/R)$ is thus expressible as proportional to $(\partial p/\partial t)_V$. The proportionality involves invariant parameters, $\rho_{vac}$, $\bar{\lambda}$ and $a$ which are characteristics of the quantized cells. $(\partial p/\partial t)_V$ is thus represented as, $(\partial p/\partial t)_V \equiv \dot{p} = k_0(\dddot{a}/a)$, where $p(t) = k_0(\ddot{a}/a)$, ($k_0$ being proportionality constant). Thus, parallel to the work performed by volume displacement in the space expansion homogeneously and isotropically across the cosmic volume, there is pressure build up throughout the volume by the jerk's presence. In a macro volume, $V(t)$, the pressure evolution can be represented with $p(t) \,\square\, k_0(\ddot{R}/R) = k_0\varepsilon' H_t^2$ with $\partial p/\partial t \approx k_0(\dddot{R}/R) = k_0\varepsilon' H_t^3$. The two stress signatures in the critically mitigated evolution thus manifest cumulatively with a ratio of $E_{\text{w}}/E_{\text{elast}} = 3\int_0^{\tau_{\text{age}}} p(t)V(t)H_t dt \div \int_0^{\tau_{\text{age}}} V(t)\dot{p}dt$. Substituting the values of pressure, $p(t) \approx k_0\varepsilon' H_t^2$, and its time derivative, $\dot{p} \approx k_0\varepsilon' H_t^3$, the stress ratio is rewritten as $E_{\text{w}}/E_{\text{elast}} \approx 3\int_0^{\tau_{\text{age}}} k_0\varepsilon' H_t^3 V(t)dt \div \int_0^{\tau_{\text{age}}} k_0\varepsilon' H_t^3 V(t)dt = 3$. As discussed in the next subsection, the estimated ratio corroborates to the vacuum field's energy characteristics. It is pertinent to mention here that the stress ratio, $E_{\text{w}}/E_{\text{elast}} \approx 3$, have some semblance with the proportions of regular (ordered) and thermal radiation power losses in the relaxation events in electrodynamics, where the maximum possible ratio in the respective relaxation rates was seen as $\dot{E}_{\text{Larmor}}/\dot{E}_{\text{Thermal}} = \pi$. The stress energy, $E_{\text{elast}}$ comes under the general category of disorderly manifested relaxation energy in H-U phenomena.

*(iii) H-U state registered in metamorphically transformed field stresses during mitigation*

In the vacuum field mitigated evolution of gravitationally active space, proportions of ordered and disordered relaxations are definable by the distinguishable energy states of the field. As per the analysis of quantized space fabrics presented in subsection 3.1.1, pure vacuum field is to be referred with the two features: (i) the basic space time fabrics of nonlocally entangled geodesics as the delocalized feature that is stretch infinitely over space having quantized energy, $2\pi c\hbar/\bar{\lambda}$, and energy density, $2\pi c\hbar/\bar{\lambda}^4 \equiv \rho_{vac}$, and (ii) the virtual lattice network formed out of the entangled geodesics, the energy density of which is based on density of states consideration of the ever present quantum oscillators with zero point energies ($\hbar\omega/2$) as can be defined within the limits of $0 < \hbar\omega \leq 2\pi c\hbar/\bar{\lambda}$. This energy density is expressible as $2\pi^2 c\hbar/\bar{\lambda}^4 \equiv \rho_{zero,\,vac}$ (say). Endowed with the two features the pure vacuum field critically mitigates on the gravitationally perturbed space in nonstationary evolution. In the mitigating interaction, the delocalized feature with density (pressure) $\rho_{vac}$ promptly reacts with the perturbed space leading to transfers of the (i) quantized lattice-cells of energy density $\rho_{zero,\,vac}$ and (ii) proportionately, the geodesics structure; their proportions follow the ratio, $\rho_{zero,vac} : \rho_{vac} \sim \pi$. The transferred cells as a component of the perturbed space occupies volumes in accommodating themselves, imparting thereby outward thrust for expansion of gravitationally active space by virtue of the state property of vacuum field as $w = -1$.

Delocalized network of nonlocally entangled geodesics that are transferred in parallel settles down with significantly additional stress of pressure-like elastic type within the gravitationally active space. The elastic stress manifests as the unseen (dark) matter with gravitational effect and it settles down without any work signature, which is unlike the repulsive field of unseen (dark) energy that registers displacement work. The elastic stress that manifests without work signature is analogue of the thermal energy in H-U phenomenon that also does not register as work energy.

With the transfer of two distinguishable quantized features of the mitigating field the classically established stress kinematics of matter and affine space curvature undergo modifications as reflected in the modeled equations (A) and (B) given in section 3.1(i) and also in the thermodynamic modeling given in 3.1(ii). In 3.1(i), the estimated dark energy proportion of 75% implies somewhat lesser content of the gravitating matters (25%; dark+normal) than the reported value (31.7%) [10]. In 3.1(ii), the indicated ratio of stress accumulations with and without work signatures in space expansion ($E_{\mathrm{w}} / E_{\mathrm{elast}} \approx 3$) usually suggests that, after considering the presence of 4.9% normal matter, the remaining 95.1% can be allotted to 71.33% dark energy and 23.77% dark matter. In this case, the total gravitating matter content of 28.77% is somewhat closer to the reported value. The low value of total matter in case 3.1(i) is possibly due to the use of somewhat lower value of the Hubble's constant in assessing the dark energy content. Now, baryonic and leptonic matters that manifest to the extent of about 5% during spontaneous symmetry breaking process in the reportedly super-cooled state (~2300 K) [11] of inflationary expansion history, are indeed an outcome of the metamorphically transformed vacuum field in its mitigation of the evolving gravitationally active space. Of the two features, the virtual lattice-cells endowed with zero point energies have the more likelihood of partially undergoing metamorphic transformation as the normal matters. The overall work energy involved in displacement action while transferring the virtual lattice-cells into the expanding space is to be accounted by the occupied volumes of the transformed (as matters) as well as untransformed cells. If matters manifest by sparing energies of virtual cells, then the apportioning of total matters and dark energy in the above mentioned estimation involving the result of subsection 3.1(ii)s will be somewhat altered. 75% of the work energy ($E_{\mathrm{w}}$) should manifest as per the metamorphic transformation as 70.1% dark energy with the 4.9% normal matter content, so that total gravitating matter gets modified to 29.9% instead of 28.75%.

Vacuum field started mitigating as soon as it encountered the super-dense state of speculated beginning of gravitationally active space with a density superseding the critically mitigating field density, $\rho_{vac}$. The mitigation involved abrupt inflationary measure with maximum possible acceleration of the field ($c^2 / \lambda'^2$) for promptly reducing the density of gravitationally perturbed state below $\rho_{vac}$. The critical mitigation at this stage is registered with a huge manifestation of dark energy and dark matter in the proportions of $\rho_{zero,vac} : \rho_{vac} \sim \pi$, wherein energy part ($\rho_{zero,vac}$) borne by virtual lattice-cells is executing the inflationary expansion; a small fraction (~5%) of this energy underwent the metamorphic transformation in the subsequent cooling course. In parallel, the delocalized energy part ($\rho_{vac}$) settled in strained state within the normal matter filled space adding thereby modified stress term in the energy-momentum equations describing the gravitating space, (refer equations A and B in 3.1(i)). The two distinguishable features of vacuum field thus run the emergent inflationary scenario.

### 3.1.1 *Nolocal features of the mitigating vacuum field*

Classically, the vacuum field is endowed with the geometric description of geodesics that define paths of free motion of test particles of infinitesimal inertia. Geodesics between a pair of space coordinates in the field, trace minimally displacing paths as are describable by the criterion of shortest segment length as, $\delta\int_{\tau_1}^{\tau_2}(dx^{\alpha}dx^{\beta}g_{\alpha\beta})^{1/2}=0$. The integrand, $(dx^{\alpha}dx^{\beta}g_{\alpha\beta})^{1/2}\equiv d\tau$ represents elemental displacement along a typical geodesic path described in between the pair of space coordinates as referred by the time-like parameter, $\tau$; $g_{\alpha\beta}(x^{\alpha})$ being the space metric. The resultant geodesics expressible by the minimization criterion as $\dot{v}_{\beta}=(\partial_{\beta}g_{\alpha\mu}v^{\alpha}v^{\mu})/2$, defines paths of free motion of test particle; $\dot{v}_{\mu}=\dot{v}^{\alpha}g_{\alpha\mu}$, $\dot{v}^{\mu}=dv^{\mu}/d\tau$ and $v^{\mu}=dx^{\mu}/d\tau$. In quantized space fabrics, however, the position and momentum of test particles obeying the uncertainty principle cannot be described by the precisely defined line traces of classically obtained geodesics. The quantum nature of space-time can be recognized by the consideration of free passage of a test particle through nonlocally mediated geodesics family $\left\{x_g^{\mu}(\tau)\right\}$ that defines the dynamic course in between a pair of arbitrarily selected space-like surfaces having the time-like separation, $\tau_2-\tau_1$. The delocalized passage will be realized by applying minimum displacement principle of linearly combined infinitesimally differed geodesics $\left\{x_g^{\mu}(\tau)\right\}$. Evolution of the mediated geodesics is thus expressible as $\delta\int_{\tau_1}^{\tau_2}\sum_g c_g^2(dx_g^{\alpha}dx_g^{\beta}g_{\alpha\beta})^{1/2}=0$, $\sum_g c_g^2=1$; $c_g^2$ representing weight factors in the linear combination are governed by canonical rules of the nonlocal mitigation of the geodesics family. The combined contribution of the geodesics helps implicating free motion of a test particle on the nonlocally entangled geodesics. The displacement criterion can be rewritten as

$$\sum_g c_g^2\delta x_g^{\beta}v_g^{\alpha}g_{\alpha\beta}(\text{at }\tau_2)-\sum_g c_g^2\delta x_g^{\beta}v_g^{\alpha}g_{\alpha\beta}(\text{at }\tau_1)+\int_{\tau_1}^{\tau_2}d\tau\sum_g c_g^2\delta x_g^{\beta}[-\dot{v}_g^{\alpha}g_{\alpha\beta}+v_g^{\alpha}v_g^{\mu}(-\partial_{\mu}g_{\alpha\beta}+\frac{1}{2}\partial_{\beta}g_{\alpha\mu})]=0,$$

where, $v_g^{\alpha}=dx_g^{\alpha}/d\tau$. The integrated terms in the above equation together can be rewritten as $(\delta x^{\beta}v_{\beta})_{\text{at }\tau_2}-(\delta x^{\beta}v_{\beta})_{\text{at }\tau_1}=C$ (say), where, $\delta x^{\beta}\Leftrightarrow\left\{\delta x_{g'}^{\beta}\right\}$ with $\delta x^{\beta}=\sum_{g'}c_{g'}\delta x_{g'}^{\beta}$, $v^{\mu}\Leftrightarrow\left\{v_g^{\mu}\right\}$ with $v^{\mu}=\sum_g c_g v_g^{\mu}$, and $v_{\beta}=v^{\alpha}g_{\alpha\beta}$, ($c_g c_{g'}=\delta_{gg'}$, Kronecker delta). Now, geodesics to be defined with parallel transport property of displacement, $dx^{\beta}\Leftrightarrow\left\{dx_{g'}^{\beta}\right\}$, that is, $v_{\beta,\text{ at }\tau_2}\equiv v_{\beta,\text{ at }\tau_1}$, (for all $\tau$), the parameter $C$ can be rewritten as $[(\delta x^{\beta})_{\text{at }\tau_2}-(\delta x^{\beta})_{\text{at }\tau_1}]v_{\beta}=C$. Geodesic family $\left\{x_g^{\mu}(\tau)\right\}$ to be described irrespective of the aribitrarily selected boundaries, parameter $C$ must be independent of $\tau_1$ or $\tau_2$. The only possible way under the arbitrariness of the variations is by the 4-orthogonal connections of $(\delta x^{\beta}v_{\beta})=0$ irrespective of $\tau$ and this in other words imply null value of $C$. Thus, mediated by the orthogonal connection, nonlocally entangled geodesics family $\left\{x_g^{\mu}(\tau)\right\}$ is describable as $\dot{v}_{\beta}=(\partial_{\beta}g_{\alpha\mu}v^{\alpha}v^{\mu})/2$. The orthogonal connection can be rewritten by using space-like unit 4-vectors $i^\beta \Leftrightarrow \left\{i_g^\beta\right\}$ of the infinitesimal variations, $\delta x^\beta \Leftrightarrow \left\{\delta x_g^\beta\right\}$ as $(i^\beta v_\beta)=0$, where, $i^\beta i_\beta = -1$. The 4-orthogonal connection representing coherent evolution of the geodesics family, $\left\{x_g^\mu(\tau)\right\}$, can be expressed in the spectral space ($k$) as $(i_k^\beta v_{\beta,k})=0$, ($k^\beta \equiv [\mathrm{k}^0, \bar{\mathrm{k}}]$ being spectral space coordinates). One can rewrite the 4-orthogonal connection by replacing the nonlocal mediator, $i_k^\beta \equiv \mathrm{i}_\mathrm{k}^0[1, \bar{\mathrm{i}}_\mathrm{k}/\mathrm{i}_\mathrm{k}^0]$ with the 4-vector, $w_k^\beta \equiv \lambda_\mathrm{k}[1, \bar{\mathrm{w}}_\mathrm{k}/c]$, that involves phase velocity, $\bar{\mathrm{w}}_\mathrm{k}/c = \omega_\mathrm{k}/c\bar{\mathrm{k}}$, ($\omega_\mathrm{k} = c\mathrm{k}^0$, $\lambda_\mathrm{k}$ being a parameter) as $(w_k^\beta v_{\beta,k})=0$. This connection represents nonlocal mediation characteristics of the evolving geodesics family in the spectral space, the 4-coordinates of which maintain perennial 4-orthogonality as $w_{\beta,k}k^\beta = 0$. Therefore, on can infer that the geodesics evolution corroborates to the proportionality, $v_k^\mu \propto k^\mu$. It amounts to the constant ratio of $\sum_k \sum_{k'} c_k c_{k'} v_k^\mu v_{k'}^\nu g_{\mu\nu} / \sum_k \sum_{k'} c_k c_{k'} k^\mu k'^\mu g_{\mu\nu} = \text{constant}$. Noting that the numerator of the ratio is unit length of 4-displacement, and also noting the equality $\sum_k c_k^2 = 1$, the ratio corroborates to $\sum_k c_k^2 k^\alpha k_\alpha = (2\pi/\bar{\lambda})^2 \equiv (2\pi/\bar{\lambda})^2 \sum_k c_k^2$, $\bar{\lambda}$ being a constant having dimension of length. $\bar{\lambda}$ represents fundamental length of the quantized space fabrics that supports the dispersion characteristics of the nonlocally mediated geodesics as $c^2 k^\alpha k_\alpha \equiv \omega_\mathrm{k}^2 - c^2\mathrm{k}^2 = (2\pi c/\bar{\lambda})^2$. The characteristics endorses involvement of the nonlocal mediator with the quantized energy of $2\pi c\hbar/\bar{\lambda} \equiv \hbar\omega$; energy equivalent mass being $3.86\times10^{-8}$ kg. The energy density of the geodesics network is therefore expressible as $\rho_{vac} = 2\pi c\hbar/\bar{\lambda}^4$ ~$1\times10^{112}$ $\mathrm{Jm}^{-3}$.

_Field mitigation characteristics in disentanglement of quantum states_

Geodesic network maintains coherence unless perturbation is extraordinarily intense leading to the critical power loss of the order of $(c/\bar{\lambda})\hbar\omega = 2\pi c^2\hbar/\bar{\lambda}^2 \sim 1\times10^{52}$ W. Coherent mediation in the geodesics network is exceptionally strong not to get perturbed but keep transmitting phase information across the world pertaining to whether the quantum mechanically entangled state involving electromagnetic, nuclear or weakly interacting members critically decohere or not by local interceptions, say, with observational probes. Undeterred mediation characteristics of the network, that establishes instant phase communication of the probing point of massive apparatus to the rest of world, instantly transmits information across no soon it receives interaction thrust with critically perturbed state of one of the entangled members.

_Mitigation characteristics of quantized space fabrics_

As discussed already, stress kinematics of the gravitating space is expressible in Friedmann-Lemaitre-Robertson-Walker space as $(\dot{R}/R)^2 = (8\pi G/3c^2)(\rho_M + \rho_{DE}) - c^2/R^2$. As compared to the mitigating vacuum field, the relative densities of (dark) energy and matters (normal + dark) can be expressed as $(\rho_M + \rho_{DE})/\rho_{vac} = \lambda'^2[1/R^2 + (\dot{R}/R)^2/c^2]$, where, $\bar{\lambda}' = (3/4\pi)^{1/6}(4\pi G\hbar/c^3)^{1/2}$. With the particular use of a coordinate frame that instantly commoves with the recession speed, the kinematic feature of space expansion, $(\dot{R}/R)^2$ will be a null, and the relative density ($\mathcal{P}$) noted in such frame will be governed by 3-curvature of the gravitating space as $\mathcal{P} \equiv (\rho_M + \rho_{DE})/\rho_{vac} = \lambda'^2/R^2$. Noting that $\lambda'^2/R^2$ can be expressed in terms of quantum cell densities of the space fabrics as $\bar{\lambda}'^2/R^2 = Q_R/Q_V$, where $Q_R = R/\bar{\lambda}'$ and $Q_V = R^3/\bar{\lambda}'^3$, ($Q_R$ and $Q_V$ being radial and volume densities respectively), the occupation probability, ($\mathcal{P}/Q_R$), can be expressed as $(\mathcal{P}/Q_R) = 1/Q_V$. The probability pertains to occupations of gravitationally active matter and energy in the quantum cells available within classically represented spherical envelope, $V = 4\pi R^3/3$, attributed by the static curvature of affine 3-space. Probabilistically one full cell out of the total numbers $Q_V$ is allotted for the active matter and energy, while the remaining $Q_V - 1$ cells are fully occupied by the vacuum field itself. Thus, the heuristically represented energy balance of matter/energy with affine space curvature given by the general relativity implicated this statistics as obeyed by the mitigating quantized space fabrics. The case, where $Q_V - 1$ is a null, the matter density approaches to that of the vacuum field, which is the uppermost limit for coherent containment in gravitationally compacted state under the field mitigation. It can be shown by using nonlocal transport characteristics that super-heavy object with coherent accumulation of baryonic mass to the extent of about $1\text{x}10^{54}$kg is possible before encountering failure of field mitigated containment in the coherent state (refer Appendix).

**4.0 H-U phenomenon as forerunner of megascopic events**

In the light of presented relaxation characteristics of accelerated charged particle one finds that the H-U predicted thermal power manifestation, proportional to (acceleration)$^4$, is indeed a reality. According to the noninertial observer, virtual pairs in his surrounding field are relatively in accelerated state because of which he notes H-U phenomenon in the surrounding vacuum. Finite expectation value of acceleration leading to the thermal relaxation under the phenomenon speaks for the criticality of perturbation of coherently evolving systems with an external field. Evolutions are known to be critically perturbed with an interacting probe in quantum state measurement, with an incoherent boundary/defect site, or under field fluctuation. Criticality is attained when perturbation supersedes the limiting nonlocal defense of mitigating field at a given acceleration. As noted already, pure vacuum field reacts to gravitational perturbation and its energy density limits the highest possible mitigation thrust according to the field acceleration, $c^2/\bar{\lambda}'$ as $(8\pi G/3c^2)\rho_{vac} = c^2/\bar{\lambda}'^2$. In a case, where the perturbation is as strong as the field so that the mitigation involves the highest nonlocal thrust, a hypothetical observer from his noninertial frame commoving with the acceleration of $c^2/\bar{\lambda}'$ will note, according to the H-U prediction, an ultrahigh thermal state of surrounding space of vacuum field. Thermally distributed stress borne by the space fabrics corroborates to the predicted temperature of about $4\text{x}10^{31}$K.

A lab frame observer will analyze the ultrahigh thermal state as disordered part relaxation, which resulted from stress yield under highest mitigating nonlocal thrust against critically high perturbation. Critically mitigating counteraction as such result in relaxations of both ordered and disordered kinds. The observer will reconcile the situation with scenarios of stress manifestations with and without work signatures. Considering energy-momentum aspects he will surmise that the perturbed part of flat space of vacuum field bearing the stress from ultrahigh thermal state of H-U phenomenon will undergo inflationary expansion as a newly emerging gravitationally active space having finite curvature with distinguishable metric. Displacement work involved in the mitigated inflationary course will be accomplished by repulsive property of cosmic energy that manifests under the mitigation. And this ordered form of stress accumulation in space will account dark energy content of the emerging space at an instant. Besides the displacement work in the inflation, H-U effected thermal stress is partly transferred as nondisplacement type energy that accumulates in parallel as stressed curvature of the emerging space signifying its gravitational potential. According to the mitigation characteristics, as discussed already, the two types of stresses accumulate in definite proportions. The observer will surmise that the emerging space constitutes mesomorphically transformed state of the thermally stressed space fabrics, and the space will have gravitating features. He will infer that the mitigating counteraction of vacuum field sustained in its incessant effort to restore coherent state with null expectation value of acceleration is responsible for the spontaneous changeover of the perturbed part of space with the abrupt expansion and consequent thermal relaxation with temperature lowering. The mitigation accomplishes the inflationary expansion with a partial loss of coherence at the minimum expense of the relaxation energy. Thermodynamic spontaneity ingrained in the transformation leads to emergence of the multi-component states of sub-atomics of matters and antimatters at some stage of super cooled state in the expansion course. According to reported information super-cooled state at $10^{23}$K and below is conducive for budding of matters/antimatters through the spontaneous symmetry breaking process [11].

The observer will further surmise that the omnipresent and incessant field mitigation sorts out matters and antimatters engaging them to distinguishably evolve in their respective spaces. Matters that evolve in his space according to cosmologically controlled expansion characteristics could be described following energy-momentum conservation in the local space. However, in terms of global energy-momentum conservations, evolutions of matters and antimatters in their respective spaces are interdependent, and for this reason the evolution characteristics of matters' space implicitly considers constraints of the partial description that antimatters as it were remain completely dispersed and frozen in a distinguishable and well mitigated state in space (negative sea). Under the vacuum field mitigation, matters unlike the state of antimatters evolve in segregated and structured forms as galaxies and clusters of galaxies where the unseen gravitational feature (dark matter) helps the structure formation and the unseen cosmic force due to dark energy avoids gravitational collapse.

As seen in externally driven perturbation, one expects that the perturbing agency bears the energy expense involved in relaxation under decoherence. However, in the particular case, where the coherently evolving vacuum field has to counteract with its highest possible defending reaction to win over an accidentally encountered perturbing 'knot' within its evolution space, the relaxation energies is to be incurred by the field itself. The 'knot' representing defect site or fluctuation derived solitary wave can locally perturb the field evolution but cannot supply energy in sustaining the relaxation processes involved. In view of this, the lab-observer will address energy-momentum conservation issue of the formed space by invoking nature's mitigation characteristics of detailed stress balance due to gravitational matter/energy and curvature of the affine space itself as is well recognized in general relativity. Energy-momentum balance includes accumulated stresses of both displacement and non-displacement types as are described in its simplistic forms by Eq.(A) and (B), section 3.1(i). Stressed geodesics and inflationary space expansion respectively account for the two types of energies of the emerging gravitationally active space.

Question however persists on the energy measure of the manifested gravitationally active space. Mass of the universe is after all finite within the observable cosmic sphere with its 13.8 billion light years evolution history. The finite mass speaks for unexplored characteristics of the field mitigation in the inflationary manifestation of the gravitating space. Coherently evolving vacuum field under critical perturbation is generally expected to part out a quantized chunk of active space fabrics with the ultrahigh gravitating energy. The observable $1x10^{53}$kg mass to be realized requires parting out of approximately $3x10^{60}$ quantum cells of the space fabrics; each cell mass being $m_p = 2\pi\hbar / c\lambda\!\!\!^- = 3.8588 \times 10^{-8}$ kg. It remains unanswered as to how to qualify a chunk of quantized space fabric which underwent the transformation in the field mitigation. As such, if the inflationary expansion begins with a minimum space required for Higgs-like bosonic phase to manifest within the thermally stress field then noting that minimum core diameter for the manifestation is $2\pi\hbar / \mu_b c \sim 1x10^{-17}$m, approximately $5x10^{-51}m^3$ volume of the vacuum field with about $1x10^{61}$J energy, (equivalently, $\sim 1x10^{44}$kg mass) is involved. The indicated space volume based on uncertainty principle is a minimum most requirement for manifestation of Higgs bosons; ultrahigh thermal state arising from H-U effect may activate a larger space volume in the symmetry breaking of the scalar field as required for the manifestation of still higher mass. Nevertheless, it is interesting to note that the energy density of the minimum activated volume is just an order of magnitude below the critical mitigation density of $\rho_{vac} \sim 1.8x10^{112}Jm^{-3}$. It is a matter of investigation as to whether the mitigating Planck field in reality commands infantry of Higgs-Bosons to make maiden entry through the doorstep of a new space mooted by Hawking-Unruh's ultrahigh thermal state. The perturbatively manifested thermal state is nevertheless conducive for the births of new space and consequently, the legal minors, namely, matters/antimatters through the symmetry breaking process under the Higgs mechanism. With the present knowledge it is not possible to quantify the inflationary order of the gravitationally active space. However, an analysis backwards can be worked out for rationalizing the uppermost limit of the manifested matters/energies with ultrahigh density by considering the maximum possible compaction of the fermionic species (baryons and leptons) in a coherent gravitational core before the core can attain criticality in the containment. This analysis is presented in Appendix-1. It is shown that nonlocal transport can help assembling the fermions in a super heavy composite core of total mass of about $10^{54}$kg having centrally denser core of neutrons ($\sim 10^{44}$kg) and a larger over-layer made of coherently packed neutron precursors.

In fact, the stated backward analysis for understanding the manifestation of finite but huge mass of newly emerging gravitationally active space out of the critically perturbed flat space, suggests for a possibility of future rebirth of universe via gravitation crunch of a super-heavy core. Irrespective of whether, there is aeonic possibility of recurrence of the birth through gravity crunch process or not, the hypothetically presumed first ever encounter of vacuum field with perturbing 'knot' remains as much an open issue as is the case of the conventionally presumed beginning from a singularity in accidental encounter with infinitely large perturbation out of the field fluctuation. Nevertheless, unlike the infinitely large field fluctuation, the perturbing 'knot' has the measure that it could critically lead to decoherence of the mitigating field of finite energy density ($\rho_{vac}$) $\sim 1x10^{112}Jm^{-3}$ and it makes the beginning of gravitationally active space from a finite dimension out of the H-U effected zone thereby avoiding singularity. Further, the well mitigated emerging space involves total energy density always less than that of the mitigating field, that is, $(\rho_{DE} + \rho_M) < \rho_{DE}^0 \equiv \rho_{vac}$, and as the result this emerged space is unable to generate yet another new one in the expansive evolution until $\rho_M$ approaches null value and the dark energy content of this space coherently merges with the mitigating field.

*Appendix*: Coherent mass accumulation and growth till attainment of critical instability in a super heavy core - H-U state manifestation in criticality.

According to the cosmological model, that considers the net effect of inward attraction of gravitating matters against the repulsive cosmic thrust, shows that segregation followed by formation of structured bodies requires normal matter content in excess of 84%. These heavenly structures constituting the observed galaxies and clusters of galaxies generally evolve with redistribution of matters by concentrating mass at gravitationally preferred locations in them. In that course they can encounter with highly energetic internal evolution processes in the dense cores under intense compressive thrust of gravity. For example, giant stars within galaxies can undergo internal thermonuclear combustion to result in the supernova events, which are observationally supported cosmic phenomena. Neutron stars or black holes are considered as possible remnants of these events. Neutron stars produced this way attain the Tolman-Oppenheimer-Volkhov limit for neutron degeneracy pressure and correspondingly, the mass limit. Generally, it is considered that with the mass accumulation beyond $2.16M_{\odot}$, ($M_{\odot}$ being mass of the sun), the degeneracy pressure cannot withstand inward pull of gravity and collapses to a black hole or to an exotic object like quark-star. Smallest observed stellar black hole is having the mass of about $5M_{\odot}$. The black hole M87 of mass, ~$1.3\times10^{40}$ kg, located in Messier galaxy (~ 53 million light years away) [12] and more recently reported supermassive black hole of mass $3.6\times10^{40}$kg in OJ 287 Galaxy (3.5 billion light yrs away) [13] are such examples. The giant clusters with supernova remnants evolve in their respective kinematic locations of the cosmic sphere. Important aspect of their evolution is the propensity of mass augmentation by the intense gravitational pull of the remnant cores with very high mass density. In binary couplings of neutron stars or black-holes, as observed in recent times, lead to the direct mass transport from one to the other core that attains higher mass in the coupling course. Mass augmentation by such direct mass transport is limited by a significant energy loss. It is also limited by the availability of other heavenly bodies in vicinity so that the attractive pull can be accomplished.

For an inertial observer, however, the transport across event horizon of a black hole is obscured by the relativistically assigned infinitely large time dilation as matter approaches thereto with signal speed. Materials including radiation, that are gravitationally attracted towards black hole stay either in the rotating accretion disc embracing the event horizon or as the stationary energy at the horizon. The inertial observer may at best describe the state of matter/energy at event horizon as a kind of standing (stationary) energy molasses. The arrest of time hinders him devising any flow process of matter/energy through event horizon inwards. He rather concludes that an energy quantum ($\hbar\omega$) though pulled inward by gravity, the pulling effect is counterbalanced by the centrifugal force of $\hbar\omega / R_{eh}$, $R_{eh} = GM / c^2$, ($R_{eh}$ being radius of event horizon of the superheavy remnant core of mass $M$ ).

Now, under the strong gravitational pull the total mass content within event horizon of a black hole is expected to segregate to a more compact volume than remaining distributed over the entire space enveloped by the horizon. Reported core density (>$10^{17}$kgm$^{-3}$) of a neutron star is, for example, much higher than averaged density taken over the volume enveloped by the event horizon. For visualizing the effect of gravitational pull inside event horizon, one may consider a changeover from the usage of standard relativistic reference of constant signal speed that freezes dynamic processes at the event horizon. Consideration can be made of the universally present reference state of nonlocal mediator, the vacuum field. The field can help augmenting mass accumulation at the central part of the overall core by sparing some energy quanta and simultaneously makeup for the loss by nonlocally transporting energy accumulated at and outside the event horizon. With such consideration one can presume that mighty vacuum field with its significantly higher energy density than that of the gravitational core can mediate in transporting energy inside the superheavy object from the horizon to the centrally dense part of the core. Energy equivalent mass accumulation increases the gravitational compression progressively with more energy transport. In the nonlocal energy transport the kinematics does not involve relativistic concept of time; instead one can refer Newtonian concept for the nonlocal way of mass augmentation of the central core.

However, in nonlocal transport, vacuum field shares its energy with the central part of the core according to the mitigating interaction probability, which is characteristically proportional to the local acceleration expressible as, $4\pi G\rho_{core}/3c^2 \sim 3.1\text{x}10^{-27}\rho_{core}$ s$^{-2}$, ($G$, the gravitational constant). For low core density ($\rho_{core}$), the acceleration magnitude is very smal as compared with the mitigating ability of vacuum field, namely, with the acceleration of $\omega_{vac}^2 \square c^2/\bar{\lambda}^2 \sim 2.8\times10^{85}$ s$^{-2}$. Nevertheless, with the slow but progressive mass accumulation and densification as $\rho_{core}$ approaches $\rho_{vac} \equiv 2\pi c\hbar/\bar{\lambda}^4$ (~1.85x10$^{112}$ Jm$^{-3}$) the energy transport through nonlocal means will be more and more efficient. Mass augmentation by coherent process is smoother as compared to the binary coupling of heavy cores, which is violent with significant dissipative losses. Direct transport through the coupling process being fast is essential at the initial augmentation of mass because of negligible efficiency of coherent transfer at low core density. Generally, the nonlocal transport having minimum dissipation loss makes up for the additional mass accumulation over the direct means but it takes aeonic long period to attain the required gravity effect for compressing the core towards the critically high density ($\rho_{core} \sim \rho_{vac}$).

*Theoretical mass accumulation and its critical limit* :

State property like mass of a super heavy gravity core can in principle be envisaged independent of the paths when alternative ways are possible. On this basis, the maximum possible mass augmentation of a core can be worked out using the path of nonlocal transport. With the progressive mass augmentation of the central part of the core there is densification in parallel. Under the compressive thrust of gravity field, the core in its densification course can encounter phase transformation with abrupt rise in mass density. Significant rise in density occurs when the initially consolidated fermionic mass of a neutron star undergoes transformation into a bosonic phase in which the repulsive thrust of degenerate fermions is no more present. However, exact path of fermionic to bosonic transformation is obscure so far.

Now, considering a simplified model, as described in the following two paragraphs, the high compression effect of the gravitational field against the degenerate fermion pressure is expressed. Central core when formed by coherent packing of a fermion of mass $m_f$ can be described by considering energy minimization of the degenerate fermionic state (of energy $E_f$) in the presence of the gravitational energy ($E_G$). These energies are respectively expressible [14] as $E_f = \frac{3}{40}\frac{3^{4/3}}{(4\pi^2)^{2/3}}\frac{h^2}{m_f R^2} N_f^{5/3} \equiv C_f N_f^{5/3} / R^2$ and $E_G = -\frac{3}{5}\frac{G m_f^2 N_f^2}{R} \equiv -B_f N_f^2 / R$, $N_f$ is number of fermions being considered. $C_f$ in SI unit for neutron or proton is about 7.3x10$^{-42}$ and for electron $C_f$ is about 1836 times higher (~1.36x10$^{-38}$). $B_f$-value for neutron or proton is about 1.12x10$^{-64}$ while the $B_f$ value for electron is 3.32x10$^{-71}$.

For a neutron core, the energy minimization leads to an equilibrium $R_n$ value as $R_n = 2C_n N_n^{-1/3} / B_n \sim 1.30 \times 10^{23} N_n^{-1/3}$, $N_n$ being neutron numbers (subscript, $n$ specifies neutron). For simplicity, the energy minimization assumes that no thermal energy is present in the core to impart its effect on $R_n$. Mass ($M_{n,\max}$) of the coherent neutron core being $M_{n,\max} = m_n N_n$, ($m_n$, the mass of a neutron, is about $1.67 \times 10^{-27} kg$) one writes $R_n = 1.54 \times 10^{14} M_{n,\max}^{-1/3}$. Gravitational field of the core of mass $M_{n,\max}$ corroborates to Newtonian acceleration of magnitude, $\ddot{R}_n = GM_{\mathrm{n,max}} / R_n^2$, where $G$ is gravitational constant. The mass is thus expressible in terms of the acceleration magnitude as $M_{\mathrm{n,max}} = 2.34 \times 10^{22} (\ddot{R}_n)^{3/5}$. Now, noting that the maximum possible acceleration of proton (mass, $m_{\mathrm{p}}$ ~1.67x10$^{-27}$kg) as corroborated by electrodynamics is $3m_{\mathrm{p}}c^4 / 2q^2 \sim 8.81 \times 10^{34}$ ms$^{-2}$ ($q$ being charge of proton) and also considering that neutron with similar mass as that of proton can attain the same acceleration, $\ddot{R}_n$ value can be approximated as $8.81 \times 10^{34}$ ms$^{-2}$. With such approximation the core mass, ($M_{\mathrm{n,max}}$), core radius ($R_n$), and core density ($d_n$) work out as 1.25x10$^{44}$kg, 0.31m, and 1.02x10$^{45}$kgm$^{-3}$ respectively. The core has event horizon of 9.3x10$^{13}$km. The evaluated core mass is about four orders of magnitude more massive than the recently reported mass of the black hole in the Messier 87 galaxy (M87) [12], while the core density is orders of magnitude higher than the reported maximum density of a neutron star (~5.9x10$^{17}$kgm$^{-3}$[15]). Gravitational compression with the accumulated mass is high enough to attain such a high density where neutron may undergo phase transformation to its quark constituents, or, to some exotic boson phase. A boson of mass greater than or equal to that of Higgs boson ($\mu_b$) needs much lower space dimension than $R_n$. For Higgs' bosons the minimum core diameter is $2\pi\hbar / \mu_b c \sim$ 1x10$^{-17}$m, and accordingly, a Higgs core of mass $M_{\mathrm{n,max}}$ would attend a density of 3x10$^{94}$kgm$^{-3}$ which is quite close to $\rho_{vac}$ ~2x10$^{95}$kgm$^{-3}$. The critical mitigation density, $\rho_{vac}$ could have been achieved by about 6.7 time increase of $M_{\mathrm{n,max}}$ value. As such, the estimated neutron core density $d_n$ is orders of magnitude lower than the critical value for attaining instability of gravitational containment under the vacuum field mitigation. In order to encounter instability the fermionic core must undergo transformation to a bosonic one, such as Higgs core with the stated mass augmentation by additional uptake of baryons and leptons in a coherent state as discussed below.

Instead of a neutron star, one may consider a coherent assembly of neutrons ($n$), and its precursors, namely, protons ($p$), electrons ($e$) and electron-antineutrinos ($\overline{\nu}_e$) with their approximate equal populations namely, $N'_{\overline{\nu}_e} \approx N'_e \approx N'_p \neq N'_n$ ($N'_{\overline{\nu}_e}$ $N'_e$, $N'_p$, and $N'_n$ being the numbers of the respective species). The considered composition ideally represents the state of neutron-core composition following a supernova event. For the assembly with the total energy of degenerate fermions as $E_f = \sum_f C_f N_f^{5/3} / R'^2$ and gravitational energy of $E_G = -\frac{3GM'^2}{5R'}$, the equilibrium $R'$ of the formed core in the absence of thermal energy can be represented in an ideal form as $R' = 2[(C_{\bar{\nu}} N_{\bar{\nu}_e}'^{5/3} + C_e N_e'^{5/3} + C_p N_p'^{5/3}) + C_n N_n'^{5/3}] / (3GM'^2/5)$, where $M'$ is the overall core mass; $M' = (N'_{\bar{\nu}_e} m_{\bar{\nu}} + N'_e m_e + N'_p m_p + N'_n m_n)$, $m_{\bar{\nu}_e}$, $m_e, m_p$ and $m_n$ respectively being the masses of antineutrino, electron, proton and neutron. Because of the large differences in the masses, $m_{\bar{\nu}_e} << m_e << m_p \approx m_n$, and therefore in the relative values of the $C_i$-coefficients as $C_{\bar{\nu}_e} >> C_e >> C_p \approx C_n$, numerator of the equilibrium $R'$ expression can be approximated as $2[C_{\bar{\nu}_e} N_{\bar{\nu}}'^{5/3} + C_n N_n'^{5/3}]$ with $N'_{\bar{\nu}_e} = N'_p$. Considering that total neutron and proton contents of the core represent practically the total mass ($M'$), radius $R'$ can be rewritten as $R' \sim 1.54 \times 10^{14} M'^{-1/3} [\varsigma(1-f')^{5/3} + f'^{5/3}]$, where, $f' \equiv N'_n / (N'_p + N'_n)$ is the fractional population of neutrons and $\varsigma = (C_{\bar{\nu}_e} / C_n) \equiv m_n / m_{\bar{\nu}_e} \sim 8.55 \times 10^8$. The $\varsigma$ value is chosen according to the recently reported antineutrino's rest energy data of 1.1 eV [16]. The $R'$ expression when considered with the gravitational field of the core having the magnitude of acceleration as $\ddot{R}' = GM'/R'^2 =$ $8.81 \times 10^{34}$ $ms^{-2}$, leads to the value of core mass as $M' = 1.25 \times 10^{44} [\varsigma(1-f')^{5/3} + f'^{5/3}]^{6/5}$.

When neutron and proton populations in the core are approximately equal as in a typical atomic nucleus, then the coherent mass works out as $M'$=1.53x$10^{54}$kg and radius, $R'$~35km (density ~8.5x$10^{39}$kgm$^{-3}$). However, when neutron population significantly dominates over its precursors, then the core mass is that of the pure neutron core as described beforehand. On the contrary, when neutrons are significantly low so that the core is essentially made of the neutron precursors, then the core mass and dimension are respectively $M' = 6.53 \times 10^{54}$ kg, $R' \sim$ 71km (density ~4.3x$10^{39}$kgm$^{-3}$). This last case would happen when a significant part of the low neutron content could form a separate phase of higher density like the one as shown in the formation of a pure neutron core of mass 1.25x$10^{44}$kg ($f'$ ~1.9x$10^{-11}$) with density value of 1.02x$10^{45}$kgm$^{-3}$ as described already. It appears that the coherent mass consolidation through the nonlocal transport can lead to formation of a composite core having the centrally located denser phase of neutrons or its transformed successor of an ultra-dense bosonic phase, and the huge but less dense over-layer of the neutron precursors of mass $M' \sim 6.5 \times 10^{54}$ kg. Such a fermionic core can encounter instability of the gravitational containment only if the dense central part could transform into bosonic phase/s with energy density approaching to that of the vacuum field. As indicated already in the case of a coherently packed neutron core of highest possible density, the transformation to Higgs-like boson makes the density quite close to the critical mitigation limit of the vacuum field. In the composite core, the huge over-layer of the neutron precursor can be sucked by the black hole-like central part in its transformed bosonic state and this pushes up the core density to higher than the mitigating density to result in inflationary space expansion the subsequent course of which will be identical to that described in section 4.0.